\documentstyle[sprocl,epsfig]{article}

\def\PLB{{\em Phys. Lett.}  B}  
  
\def\PRD{{\em Phys. Rev.} D}  
  
\def\EPJC{{\em Eur. Phys. J.} C}  
  
\def\be{\begin{equation}}  
\def\ee{\end{equation}}  
\def\bea{\begin{eqnarray}}  
\def\eea{\end{eqnarray}}  
  
\begin{document}  
  
\title{HIGGS OR DIJET PRODUCTION IN DOUBLE RAPIDITY GAP EVENTS\footnote{Presented at the 
8th International Workshop on Deep Inelastic Scattering, Liverpool, 25--30 April 2000}}  
  
\author{V. A. KHOZE, A. D. MARTIN, M. G. RYSKIN}  
  
\address{Department of Physics, University of Durham, DH1 3LE, England}   
  
  
\maketitle\abstracts{We quantify the rate of double diffractive Higgs/dijet production at $pp$  
(or $p\bar{p}$) colliders.  The suppression due to QCD bremsstrahlung is calculated  
perturbatively up to single $\log$ accuracy.  The survival probability of the rapidity gaps is  
discussed.  A first comparison with experiment is made.}  
  
At first sight the process $p + p \rightarrow p +$ gap $+H +$ gap $+ p$ looks to be a  
promising way to search for an intermediate mass Higgs boson.  For such events the signal to  
background ratio is much better for the $H \rightarrow b\bar{b}$ channel since the  
background $b\bar{b}$ dijet rate is suppressed in double rapidity gap events due (i) to the  
absence of the colour octet $b\bar{b}$ state and (ii) to the polarization structure of double  
diffractive $b\bar{b}$ production (which has the same selection rules as $\gamma\gamma  
\rightarrow q\bar{q}$ process in the $J_z = 0$ channel, where the LO contribution 
vanishes \cite{FKM}).  Moreover here we  
have much better $b\bar{b}$ mass resolution than in an inelastic event where the presence of  
a large multiplicity of secondary particles tends to wash out the Higgs peak.  On the other  
hand the cross section for the signal (and the background) are considerably suppressed  
\cite{KMRH,KMR} by the small survival probability of rapidity gaps, $W = S^2 T^2$.  First  
there is a probability $S^2$ that the gaps are not filled by soft rescattering, that is by an  
underlying interaction and, second, factor $T^2$ is the probability not to radiate extra gluons  
from the hard subprocess.  
  
The basic mechanism for the process is shown in Fig.~1, where it turns out that the typical  
values of $Q_t$ of the gluon, which screens the colour, are much smaller than $M_H$ but are  
yet sufficiently large for perturbative QCD to be applicable.  The corresponding longitudinal  
component $x^\prime$ satisfies $x_\pm^\prime \ll x_\pm$, where light-cone fractions  
$x_\pm$ refer to the active gluons of the hard subprocess.  The amplitude, to single $\log$  
accuracy, is  
\be  
\label{eq:a1}  
{\cal M} \; = \; A \pi^3 \: \int \: \frac{d^2 Q_t}{Q_t^4} \: f_g (x_+, x_+^\prime, Q_t^2,  
M_H^2/4) \: f_g (x_-, x_-^\prime, Q_t^2, M_H^2/4)  
\ee  
where the $gg \rightarrow H$ vertex factor $A^2 = K (\sqrt{2}/9 \pi^2) G_F \alpha_S^2  
(M_H^2)$ with the NLO $K$ factor $K \simeq 1.5$.  The unintegrated gluon densities take  
the form \cite{KMR}  
\be  
\label{eq:a2}  
f_g (x, x^\prime, Q_t^2, M_H^2/4) \; = \; R_g \: \frac{\partial}{\partial \ln Q_t^2} \left [  
\sqrt{T (Q_t, M_H/2)} \: xg (x, Q_t^2) \right ]  
\ee  
where $\sqrt{T}$ arises because the survival probability is only relevant to the hard gluon.   
The multiplicative factor $R_g$ is the ratio of the skewed $x^\prime \ll x$ integrated gluon  
distribution to the conventional one, $xg (x, Q_t^2)$.  $R_g \simeq 1.2 (1.4)$ at LHC  
(Tevatron) energies.  Finally the bremsstrahlung survival probability $T^2$ is given by  
\be  
\label{eq:a3}  
T (Q_t, \mu) \; = \; \exp \left (- \: \int_{Q_t^2}^{\mu^2} \: \frac{\alpha_S (k_t^2)}{2 \pi} \:  
\frac{dk_t^2}{k_t^2} \: \int_0^{1 - k_t/\mu} \: zP_{gg}(z) dz \right ).  
\ee  
In practice we also include the quark contribution, see \cite{KMR} for this and other details.   
At first sight integral (\ref{eq:a1}) appears divergent at small $Q_t$.  However the Sudakov  
form factor $T$ strongly suppresses the infrared contribution.  The saddle points of the  
integral are located near $Q_t^2 = 3.2 (1.5)$~GeV$^2$ at LHC (Tevatron) energies.  
 
\begin{figure}[t]  
\begin{center}
\epsfig{figure=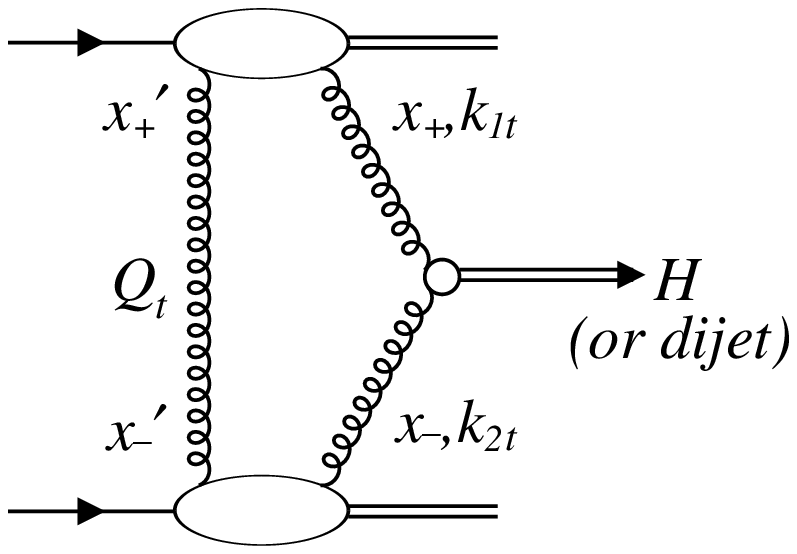,height=1.5in}  
\caption{  \label{fig:fig1}}  
\end{center}
\end{figure}  
  
It is important to emphasize that the double diffractive process does not satisfy conventional  
factorization properties.  We see amplitude (\ref{eq:a1}) satisfies $Q_t$ factorization.   
However even this is violated when soft rescattering effects are included.  
  
Amplitude (\ref{eq:a1}) is written for the exclusive process where $k_{1t} \simeq k_{2t}  
\simeq Q_t$.  The modification for the inclusive process, $pp \rightarrow X +$ gap $+ H +$  
gap $+ Y$, is given in \cite{KMR,KMRH} where it was found that the cross section is much  
larger.  
  
Our most recent calculation \cite{KMR} was performed to single $\log$ accuracy and used  
more realistic unintegrated skewed gluons, which enhances the cross section as compared to  
previous results \cite{KMRH,KMRD}. We have compensation of the factorization and  
renormalisation scale dependences between the NLO vertex $A$ and the Sudakov form factor  
$T$.  All this considerably increases the stability of our perturbative predictions of the cross  
sections, and we anticipate the higher order $\alpha_S$ effects will give, at most, about $\pm  
40\%$ uncertainty.  
  
The main uncertainty arises from the survival probability $S^2$ of the rapidity gaps with  
respect to soft rescattering effects.  This effect was extensively studied, see for example  
\cite{KMR,EML}.  It is model dependent and reflects the spatial distribution taken for gluons  
in the proton.  An optimistic estimate is $S^2 = 0.1$, which we used as the default value, but  
most probably the actual value is lower and even $S^2 = 0.01$ is not excluded \cite{KMR}.   
Assuming $S^2 = 0.1$ we find, for $m$(Higgs) = 120~GeV, that $\sigma$(exclusive) =  
5.7~fb at the LHC energy.  On the other hand for inclusive production the cross section is of  
the order of 100 (10)~fb, taking rapidity gaps $\Delta \eta = 2 (3)$.  
  
A closely related observable process is the double diffractive central production of a pair of  
high $E_T$ jets with rapidity gaps on either side of the pair.  Essentially we simply replace  
the $gg \rightarrow H$ subprocess by that for $gg \rightarrow$ dijet.  The dijet rate is much  
larger than that for Higgs and so collider experiments should be able to directly test the QCD  
estimates \cite{KMRD,KMR} and measure $S^2$.  
  
A search for double diffractive dijet events was reported \cite{CDF} at this Workshop.  The  
upper limit for such events at the 95\% CL is 3.7~nb for $E_t ({\rm jet}) > 7$~GeV.  It was  
noted that this cross section is ${\cal O} (10^3)$ smaller than the theoretical calculation  
presented\footnote{It was stressed \cite{B} that this calculation does not include the survival  
probability $S^2$.} in \cite{B}.  This theoretical model, together with those of  
refs.~\cite{BL,EML1} for Higgs production, uses a non-perturbative two-gluon Pomeron  
approach where the gluon propagator is modified so as to reproduce the total $pp$ cross  
section.  It is known that such a non-perturbative gluon normalisation will overestimate the  
double diffractive cross section \cite{KMR,B}.  On contrary, a realistic unintegrated gluon  
density, determined from conventional gluons of global parton analyses, was used in  
\cite{KMR}.  There a much smaller cross section was anticipated.  Indeed for the CDF  
kinematics we predict an exclusive dijet cross section of about 1~nb, taking $S^2 = 0.1$,  
which may be enhanced by up to a factor of 2 to allow for proton/antiproton dissociation.   
Unlike other approaches this prediction is well within the experimental limit.  
  
Another check of our perturbative approach is the behaviour of the dijet cross section with  
$E_T$ (jet).  Due to the $x$ dependence of the perturbative gluon, we predict a steeper  
$E_T$ fall off than the non-perturbative model.  Our results also show \cite{KMR} a strong  
increase of double diffractive processes with increasing energy, that again arises because of  
the growth of gluon densities with increasing $1/x \simeq s/M^2$, which is advantageous for  
the LHC.  On the contrary the predictions of the non-perturbative approaches  
\cite{BL,B,EML1} depend only weakly on energy through the energy dependence of the  
\lq\lq soft\rq\rq\ cross section which was used to normalise the two-gluon exchange  
amplitude.  
  
\section*{Acknowledgments}  
We would like to thank K. Borras, K. Goulianos and E.M. Levin for fruitful discussions.   
VAK thanks the Leverhulme Trust for a Fellowship.

\end{document}